# First-principles study of Vickers hardness and thermodynamic properties of Ti$_3$SnC$_2$ polymorphs

M. A. Rayhan[1], M. A. Ali*[,2], S. H. Naqib[1] and A. K. M. A. Islam[3]

[1]Department of Physics, Rajshahi University, Rajshahi-6205, Bangladesh
[2]Department of Physics, Chittagong University of Engineering and Technology, Chittagong-4349, Bangladesh
[3]International Islamic University Chittagong, 154/A College Road, Chittagong, Bangladesh

**Abstract**

We have investigated Vickers hardness and the thermodynamic properties of the recently discovered nanolaminate carbide Ti$_3$SnC$_2$ polymorphs using the first-principles calculations. The chemical bonding shows a combination of covalent, ionic and metallic types. The strong covalent bonding is mainly responsible for high Vickers hardness of Ti$_3$SnC$_2$ polymorphs. Thermodynamic properties are studied using the quasi-harmonic Debye model. The variation of bulk modulus, thermal expansion co-efficient, specific heats, and Debye temperature with applied pressure (P) and temperature (T) are investigated systematically within the ranges of 0 - 50 GPa and 0 - 1000 K. The calculated results have been compared with available experimental and theoretical data.

## 1 Introduction:

The discovery of the layered ternary ceramics with a common formula M$_{n+1}$AX$_n$ (MAX) phases (with $n$ = 1, 2 or 3, M is early transition metal, A is an A-group element in the periodic table, and X is either C or N) by Nowtony *et al*. [1] have drawn attention among the research community due to their outstanding properties having characteristics of both ceramic and metal [2]. The MAX compounds exhibit remarkable physical properties, such as high mechanical strength, good electrical conductivity, exceptional shock resistance and damage tolerance, fully reversible plasticity, and high thermal conductivity [3-7]. These unique set of properties makes them potentially interesting for industrial applications.

Currently there are over 70 MAX [8]. Six members of 312 phases are: Ti$_3$SiC$_2$, Ti$_3$GeC$_2$, Ti$_3$AlC$_2$, Ta$_3$AlC$_2$, Ti$_3$SnC$_2$ and V$_3$AlC$_2$. The first Sn-containing 312 phase, Ti$_3$SnC$_2$, was discovered by Dubois *et al*. [9] in 2007. The polymorphism of the MAX phases has also attracted some attention in recent years [10]. The polymorphs of Ti$_3$SnC$_2$ have been identified. Both of them crystallize in a hexagonal structure with the space group P6$_3$/mmc but have different atomic positions. One is α-Ti$_3$SnC$_2$ and the other one is β-Ti$_3$SnC$_2$. Ti$_3$SnC$_2$ polymorphs are promising materials for high temperature applications.

Earlier studies of Ti$_3$SnC$_2$ polymorphs have been reported in literature. The structure of the ternary carbide Ti$_3$SnC$_2$, based on first-principles calculations has been presented by M. B. Kanoun *et al*. [11]. M. W. Barsoum *et al*. also studied the phase stability, electronic structure, compressibility, elastic and optical properties [12]. Mechanical properties have been studied by S. Dubois *et al*. [13]. Most of the investigations dealt with structural, elastic, electronic and optical properties, but the theoretical hardness and thermodynamic properties have not been studied thoroughly. Investigation of the thermodynamic properties are important for the understanding of the specific behavior of Ti$_3$SnC$_2$ polymorphs under high pressure and high temperature environments. Therefore, we have undertaken this project to study the Vickers hardness and thermodynamic properties of Ti$_3$SnC$_2$ polymorphs for the first time.

## 2 Method of Calculation:

### 2.1 *Total energy electronic structure calculations:*

The zero-temperature energy calculations have been carried out using the CASTEP code [14] by employing pseudopotential plane-wave approach based on the density functional theory (DFT). The electronic exchange-correlation potential is evaluated under the generalized gradient approximation (GGA) with the functional developed by Perdew-Burke-Emzerhog (PBE) [15]. To describe the interaction between ion and electron, ultrasoft Vanderbilt-type pseudopotentials are employed for Ti, Sn and C atoms [16]. A plane-wave cutoff energyof 500 eV is used for all cases. For the sampling of the Brillouin zone, the Monkhorst-Pack scheme [17] is used to generate a uniform grid of k-points along the three axes in reciprocal space, and a 11x11x2 special k-points are in use to achieve geometry optimization. All the structures are relaxed by BFGS minimization technique [18]. Geometry optimization is achieved using convergence thresholds of 5x10$^{-6}$ eV/atom for the total energy, 0.01 eV/Å for the maximum force, 0.02 GPa for the maximum stress and 5x10$^{-4}$ Å for maximum displacement.

---

* *Corresponding author*: ashrafphy31@gmail.com

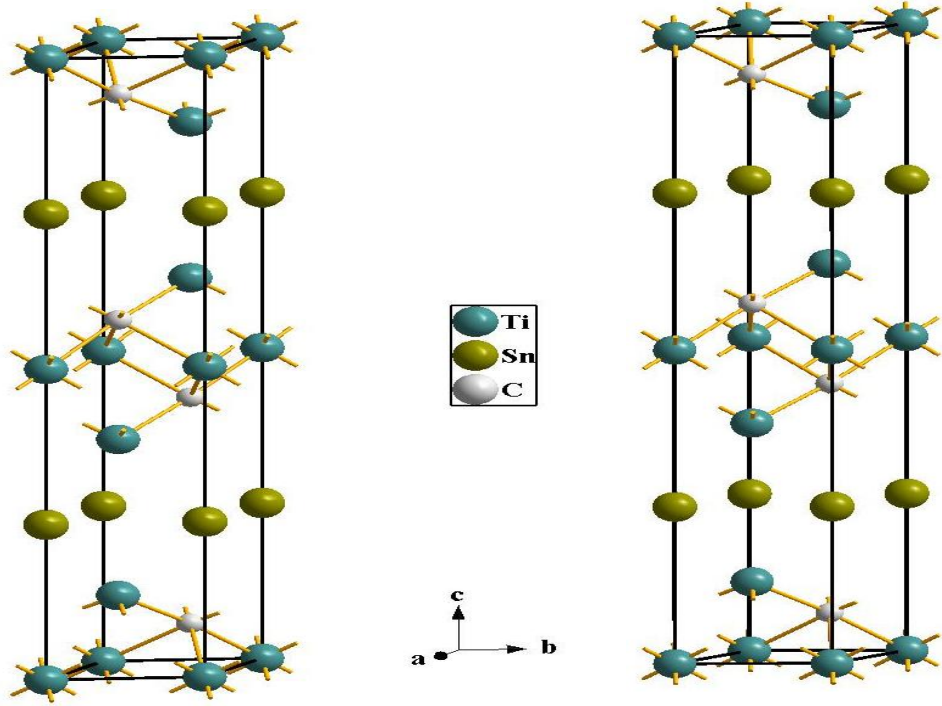

Fig. 1. The crystal structure of (a) α- Ti$_3$SnC$_2$ and (b) β-Ti$_3$SnC$_2$.

## 2.2 *Thermodynamic properties*

The two specific heats ($C_p$, $C_v$), volume thermal expansion coefficient, Debye temperature, thermal conductivity, etc., are some of the thermodynamic properties of a solids. In order to determine these thermodynamic properties we need $E(V)$ data. But we have only equilibrium energy $E_0$ and equilibrium volume $V_0$ data (in addition to zero pressure bulk modulus $B_0$ and its derivative $B_0'$ ) from DFT calculations. To proceed we use the following procedure.

The equation of state (EOS) and the chemical potential are two of the key thermodynamic properties of a solid. The EOS of a given crystalline phase determines its behavior with respect to changes in the macroscopic variables, mainly pressure (*P*) and temperature (*T*) [19]. The third-order Birch–Murnaghan isothermal equation of state is given by

$$P(V) = \frac{3B_0}{2}\left[\left(\frac{V_0}{V}\right)^{\frac{7}{2}} - \left(\frac{V_0}{V}\right)^{\frac{5}{3}}\right]\left\{1 + \frac{3}{4}\left(B_0' - 4\right)\left[\left(\frac{V_0}{V}\right)^{\frac{2}{3}} - 1\right]\right\} \quad (1)$$

*E(V)* is found to be

$$E(V) = E_0 + \frac{9V_0 B_0}{16}\left\{\left[\left(\frac{V_0}{V}\right)^{\frac{2}{3}} - 1\right]^3 B_0' + \left[\left(\frac{V_0}{V}\right)^{\frac{2}{3}} - 1\right]^2 \left[6 - 4\left(\frac{V_0}{V}\right)^{\frac{2}{3}}\right]\right\} \quad (2)$$

Given the energy of a solid (E) as a function of the molecular volume (V), the Gibbs program uses a quasi-harmonic Debye model to generate the Debye temperature, normalized volume, specific heat capacity, bulk modulus, volume thermal expansion coefficient. In the quasi harmonic Debye model the non-equilibrium Gibbs function *G(T, P)* can be written as

$$G(T, p) = E_{tot} + (E_{ZPE} - T\,S_{vib}) - T\,S_{conf} + p\,V \quad (3)$$

Here $E_{tot}$ = Total energy, which is directly obtained from the electronic structure calculations at $T = 0$ and $P = 0$. The second term is the zero point energy, and the third term vibrational entropic. The fourth term $T\,S_{conf}$ is the configurational entropy and the last one is the $pV$ term corresponds to the constant hydrostatic pressure condition. We are considering a perfect crystal whose only degrees of freedom are vibrations, and so $S = S_{vib} + S_{conf} = S_{vib}$. We may thus write

$$G(T, p) = E_{tot} + A_{vib} + p\,V, \quad \text{where } A_{vib} = E_{ZPE} - T\,S_{vib}.$$

Term $A_{vib}$ is given in Debye model as [20-24]

$$A_{vib}(\Theta;T) = nkT\left[\frac{9\Theta}{8T} + 3\ln\left(1 - e^{-\Theta/T}\right) - D\left(\Theta/T\right)\right] \quad (4)$$

where $n$ is the number of atoms per formula unit, $D(\Theta/T)$ represents the Debye integral, and for isotropic solid, $\Theta$ is expressed as [21]

$$\Theta = \frac{h}{K}\left[6\pi^2 V^{\frac{1}{2}} n\right]^{\frac{1}{3}} f(\sigma)\sqrt{\frac{B_S}{M}} \quad (5)$$

$M$ being the molecular mass per unit cell and $B_s$ the adiabatic bulk modulus. The static compressibility is given by [19]

$$B_S = B(V) = V\frac{d^2 E(V)}{dV^2} \quad (6)$$

The heat capacity $C_v$ and thermal expansion coefficient α are given as follows [25]

$$C_{V,vib} = 3nk\left[4D\left(\frac{\Theta}{T}\right) - \frac{3\Theta/T}{e^{\Theta/T} - 1}\right] \quad (7)$$

and

$$\alpha = \frac{\gamma C_V}{B_T V} \quad (8)$$

where γ is the Grüneisen parameter, defined as

$$\gamma = -\frac{d\ln\Theta(V)}{d\ln V} \quad (9)$$

3. **Results and Discussion**

3.1 *Vickers Hardness*

The Mulliken bond populations are calculated to understand the bonding behavior as well as to obtain Vickers hardness ($H_V$) of α- and β-Ti$_3$SnC$_2$ polymorphs. The relevant formula for the hardness is given as [26, 27]

$$H_V = \left[\prod_1^\mu \left\{740(P^\mu - P^{\mu'})(v_b^\mu)^{-5/3}\right\}^{n^\mu}\right]^{1/\sum n^\mu} \quad (10)$$

where $P^\mu$ is the Mulliken overlap population of the μ-type bond, $P^{\mu'} = \dfrac{n_{free}}{V}$ ($n_{free}$ is the number of free electron and $V$ is the crystal volume) is the metallic population and $V_b^\mu$ is the bond volume of μ-type bond. The coefficient 740 is due to the diamond like bond.

The calculated results are given in Table 1.1. The Mulliken bond populations gives the degree of overlap of electron clouds of the two bonding atoms. The strong covalency of the chemical bonding occurs due to the highest value of bond populations and its low value implies that the chemical bond exhibits strong ionicity. Therefore, it can be seen that the Ti1-C bonds possess stronger covalent bonding than Ti2-C bond in α-Ti$_3$SnC$_2$ phase. In case of α-Ti$_3$SnC$_2$ the population of Ti1-Ti1 bond is low. This indicates that, the Ti1-Ti1 bonds of α-Ti$_3$SnC$_2$ possess lower covalency. Similarly, for β-Ti$_3$SnC$_2$, the Ti1-C bonds possess stronger covalent bonding than Ti2-C bond. The bond of zero population does not contribute to the hardness calculation, so the hardness of α- and β-Ti$_3$SnC$_2$ is due to the hardness of Ti1-C and Ti2-C bond. Here in Ti-C bond, the electronegativity difference between Ti (1.54) and C (2.55) is 1.01 and in Ti-Ti bond the difference is exactly zero. As per this rule, the Ti-C bonds possess polar covalent bonding and Ti-Ti bonds hold non-polar covalent bonding. Polar covalent bonds are always stronger than non-polar covalent bonds. Again, the degree of metallicity may be defined as $f_m = P^{\mu'}/P^{\mu}$. In our calculations, the value of $f_m$ for Ti1-Ti1 bond in α-Ti3SnC2 phase is 0.11, which is larger than that of other bond (i.e. Ti1-C, Ti2-C: 0.011, 0.015), indicating that the Ti1-Ti1 bond is more metallic than other bonds. Ti2-C bond is more metallic than Ti1-C bond in β-Ti$_3$SnC$_2$ phase. After calculating the individual bond hardness of all bonds in the crystal the total Vickers hardness of the compound is found by taking geometric average of these bonds' hardness. The theoretically calculated values of the Vickers hardness for α- and β-Ti$_3$SnC$_2$ phase are 9.6 GPa and 4.9 GPa respectively. The experimental hardness of α-Ti$_3$SnC$_2$ is found to be 9.3 GPa [28] by the nanoindentation process with Nix and Gao model [29]. We may conclude that β-Ti$_3$SnC$_2$ phase is relatively soft and easily machinable compared to α-Ti$_3$SnC$_2$ phase. From Table 1 we observe that, the calculated Vickers hardness of α-Ti$_3$SnC$_2$ is almost two times greater than that of β-Ti$_3$SnC$_2$ which is expected.

Table 1. Calculated Mulliken bond overlap population of $\mu$-type bond $P^{\mu}$, bond length $d^{\mu}$, metallic population $P^{\mu'}$, bond volume $V_b^{\mu}$ and Vickers hardness of $\mu$-type bond $H_V^{\mu}$ and $H_v$ of α- and β-Ti$_3$SnC$_2$ polymorphs.

| Phase | Bond | $d^{\mu}$ (Å) | $P^{\mu}$ | $P^{\mu'}$ | $V_b^{\mu}$ (Å$^3$) | $H_V^{\mu}$ (GPa) | $H_V^{Th.}$(GPa) | $H_V^{Exp.}$(GPa) |
|---|---|---|---|---|---|---|---|---|
| α-Ti$_3$SnC$_2$ | Ti1-C | 2.09729 | 1.21 | 0.013 | 5.124 | 58.461 | 9.6 | 9.3 [28] |
|  | Ti2-C | 2.21927 | 0.87 | 0.013 | 6.072 | 31.075 |  |  |
|  | Ti1-Ti1 | 4.70138 | 0.12 | 0.013 | 57.73 | 0.079 |  |  |
| β-Ti$_3$SnC$_2$ | Ti1-C | 2.09803 | 1.21 | 0.074 | 18.5070 | 6.725 | 4.9 |  |
|  | Ti2-C | 2.21036 | 0.87 | 0.074 | 21.6413 | 3.534 |  |  |

[$H_V^{Th}$ Theoretical hardness, $H_V^{Exp.}$ Experimental hardness]

### 3.2 *Thermodynamic Properties*

We have investigated thermodynamic properties of Ti$_3$SnC$_2$ polymorphs by using the quasi-harmonic Debye model, detailed description of which can be found in literature [30, 31]. The thermodynamic properties are calculated in the temperature range from 0 to 1000 K, where the quasi-harmonic Debye model remains fully applicable. The pressure effect is studied in the 0 to 50 GPa range. Here we have calculated the bulk modulus, normalized volume, specific heats, thermal expansion coefficient and Debye temperature at various temperatures and pressures for the first time. For this we have made use of *E-V* data obtained from the third-order Brich-Murnaghan equation of state [32] using zero temperature and zero pressure equilibrium values, $E_0$, $V_0$, $B_0$, obtained from first principles calculations as discussed earlier. It can be seen that, there is no noticeable difference for the temperature and pressure dependence properties between the two polymorphs under study (Fig. 1).

We see that from Fig. 2(a), there is hardly any difference in the values of *B* for the two phases and these vary almost identically as a function of temperature. That means for the same compressive stress applied to both the phases at a particular temperature results in the same volume strain in both phase. At 0 K, the bulk modulus is 160 GPa for α-Ti$_3$SnC$_2$ and is 162 GPa for β-Ti$_3$SnC$_2$, which is consisted with the previous studies [11, 12]. It is seen that the bulk modulus is nearly a constant when T < 100 K. However, for T > 100 K, the bulk modulus decreases with the increasing temperature.

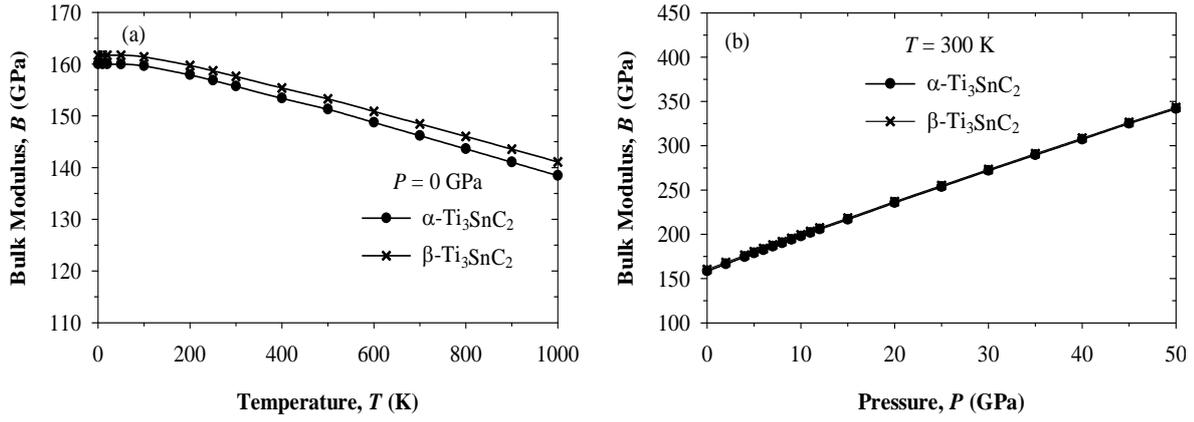

Fig. 2. (a)Temperature dependence and (b) pressure dependence of the bulk modulus for α- and β-$Ti_3SnC_2$.

The bulk modulus of α-$Ti_3SnC_2$ and β-$Ti_3SnC_2$ drops by 13.5% and 12.8%, respectively, from 0 to 1000 K. From Fig. 2(b), it is seen that bulk modulus increases with increasing pressure and the shape of curve is nearly linear for both the phases.

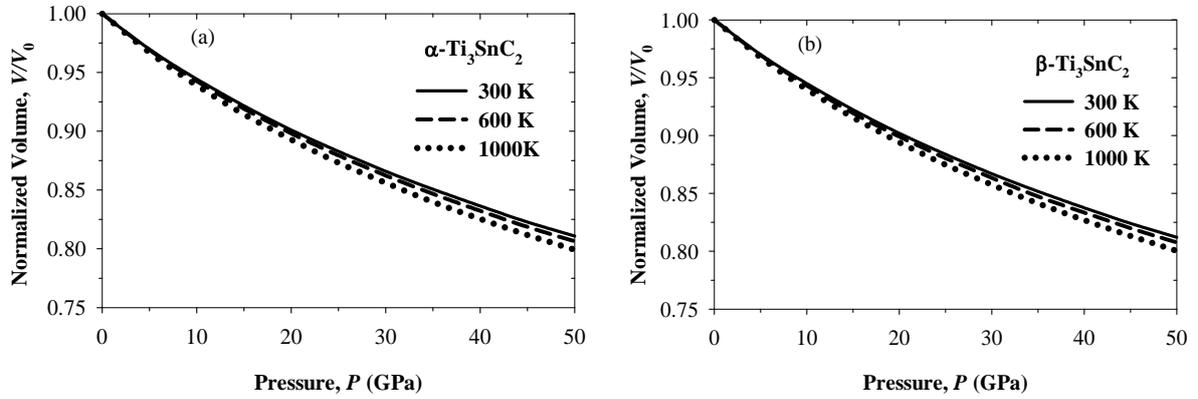

Fig. 3. Normalized volume-pressure diagram for (a) α-$Ti_3SnC_2$ and (b) β-$Ti_3SnC_2$ at different temperatures.

The pressure and temperature dependence of the relative volume $V/V_0$ of α- and β-$Ti_3SnC_2$ are shown in Fig. 3(a) and 3(b). It is seen that the unit cell volume decreases smoothly and no abrupt change occurs with increasing pressure for both the phases, indicating that the crystal structure is stable up to a pressure of 50 GPa. It can be seen that on compression, the reduction in volume for α-Ti3SnC2 is greater than β-$Ti_3SnC_2$, which is due to the higher bulk modulus of β-$Ti_3SnC_2$.

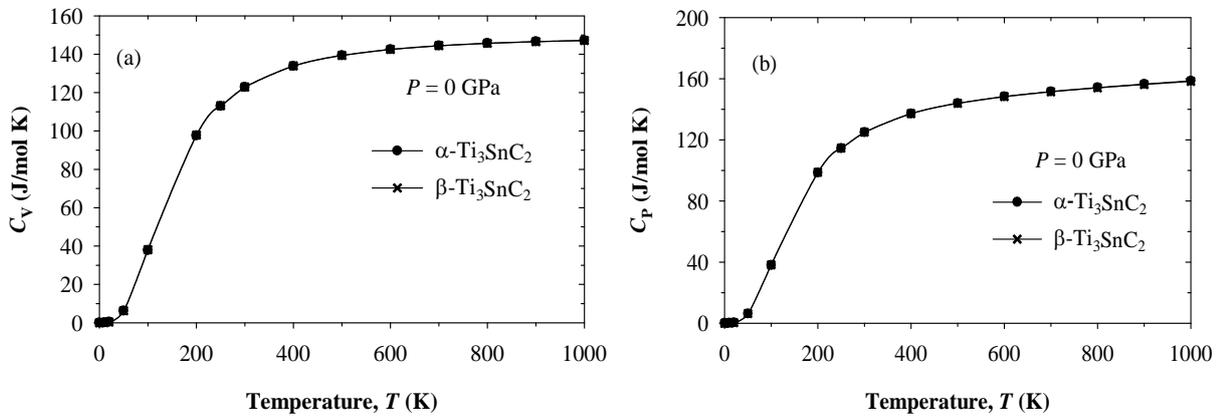

Fig. 4. The specific heat (a) at constant volume and (b) constant pressure with temperature for α- and β-$Ti_3SnC_2$ at pressure $P = 0$ GPa.

The specific heat capacity, $C_V$ at constant-volume and $C_P$ at constant-pressure for α- and β-Ti$_3$SnC$_2$ polymorphs as a function of temperature are calculated and shown in Fig. 4(a) and 4(b), respectively. The differences in $C_V$ and $C_P$ of α- and β-Ti$_3$SnC$_2$ polymorphs are minute, meaning that the effect of different M-A bonding on the specific heat is negligible. It is seen that both heat capacities increase with increasing temperature. These results indicate that phonon thermal softening occurs when the temperature is raised. In the low temperature limit, $C_V$ of α- and β-Ti$_3$SnC$_2$ phases obeys the expected Debye $T^3$ power law behavior. At high temperature (T > 400 K) it follows the Dulong and Petit law and $C_V$ approaches the classical asymptotic limit $C_V=3nNk_B$ = 149.6 J/mol K The values of $C_P$ for α- and β-Ti$_3$SnC$_2$ are slightly larger than the $C_V$, which can be explained by the relation between $C_P$ and $C_V$ as follows

$$C_P - C_V = \alpha_V^2(T)BVT \tag{11}$$

where $\alpha_V$, $B$, $V$ and $T$ are the volume thermal expansion coefficient, bulk modulus, volume and absolute temperature, respectively.

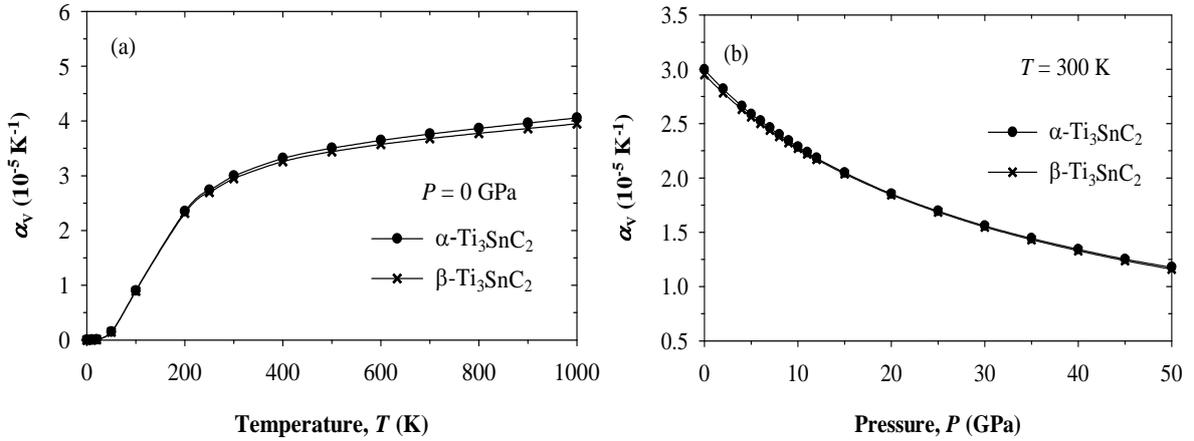

Fig. 5. The volume thermal expansion coefficient with (a) temperature and (b) pressure for α- and β-Ti$_3$SnC$_2$.

The volume thermal expansion coefficient, $\alpha_V$ as a function of temperature and pressure are shown in Fig. 5(a) and 5(b), respectively. Notice that $\alpha_V$ increases rapidly with increasing temperature at low temperature region of $T < 300$ K and increases gradually at high temperatures. The thermal expansion coefficient of α-Ti$_3$SnC$_2$ is greater than that of β-Ti$_3$SnC$_2$ phase, which is consistent with the fact that α-Ti$_3$SnC$_2$ possesses a lower bulk modulus and higher volume reduction when compressed. It is established that the volume thermal expansion coefficient is inversely related to the bulk modulus of a material. The calculated values of $\alpha_V$ at 300 K for α- and β-Ti$_3$SnC$_2$ polymorphs are $2.99\times10^{-5}$ K$^{-1}$ and $2.95\times10^{-5}$ K$^{-1}$, respectively. The estimated linear expansion coefficients ($\alpha = \alpha_V/3$) are $9.97\times10^{-6}$ K$^{-1}$ and $9.83\times10^{-6}$ K$^{-1}$, respectively.

As a fundamental parameter, the Debye temperature correlates with many physical properties of solids, such as specific heat, elastic constants, and melting temperature. At low temperatures the vibrational excitations originates solely from acoustic vibrations. Hence, at low $T$ the Debye temperature obtained from elastic constants is the same as that determined from specific heat measurements. From the elastic constants, one can get the Debye temperature ($\Theta_D$) using the following formulae [33]

$$\Theta_D = \frac{h}{k_B}\left(\frac{3n}{4\pi V_0}\right)^{\frac{1}{3}} v_a \tag{12}$$

where $h$ is Planck's constant, $k_B$ is Boltzmann's constant, $n$ is the number of atoms in unit cell, and $V_0$ is the unit cell volume. The average sound velocity $v_a$ is approximately expressed as [34]

$$v_a = \left[\frac{1}{3}\left(\frac{2}{v_t^3} + \frac{1}{v_l^3}\right)\right]^{-\frac{1}{3}} \tag{13}$$

where $v_t$ and $v_l$ are the transverse and longitudinal elastic wave velocities, respectively, which can be obtained from Navier's equation [33]

$$v_t = \sqrt{\frac{G}{\rho}}, \quad v_l = \sqrt{\frac{\left(B_S + \frac{4}{3}G\right)}{\rho}} \qquad (14)$$

where $G$ is the shear modulus, $B_S$ is the adiabatic bulk modulus and $\rho$ is the density.

Using above equations, calculated Debye temperatures were found to be 659.9 K and 494.7 K for α- and β-$Ti_3SnC_2$ polymorphs at 0 K and 0 GPa, respectively. Unfortunately, there is no availability of theoretical or experimental data for α- and β-$Ti_3SnC_2$ polymorphs.

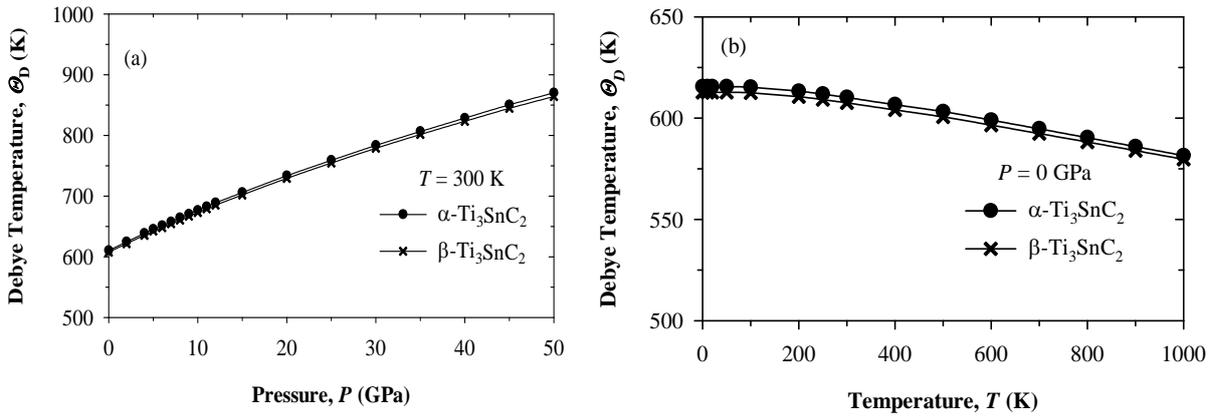

Fig. 6. The (a) pressure dependence of Debye temperature and the (b) pressure dependence of Debye temperature for α- and β-$Ti_3SnC_2$.

Fig. 6(a) shows the pressure dependence of Debye temperature $\Theta_D$ at 300 K of α- and β-$Ti_3SnC_2$ polymorphs. It is seen that the Debye temperature increases almost linearly with pressure. Fig. 6(b) displays the temperature dependence of $\Theta_D$ at P = 0 GPa. It is clear that $\Theta_D$ of α- and β-$Ti_3SnC_2$ phases remains unchanged as T < 100 K and decreases linearly as T > 100 K. Debye temperature $\Theta_D$ is related to the maximum thermal vibration frequency of a solid. The variation of $\Theta_D$ with pressure and temperature reflects the fact that the thermal vibration frequency of the particles in α- and β-$Ti_3SnC_2$ phases changes with pressure and temperature. Since vibration frequency is proportional to square root of the stiffness within the harmonic approximation, $\Theta_D$ can be used to measure the stiffness of solids [35]. Usually, a solid with high modulus and hardness will possess high Debye temperature. For example, $\Theta_D$ of diamond is 2240 K, much higher than 402 K of graphite [35].

## 4 Conclusion

To investigate the Vickers hardness and thermodynamic properties of α- and β-$Ti_3SnC_2$, we have used first-principles calculations which is based on density functional theory and the well established equation of states. Mulliken population analysis reveals that covalent bonding dominates in these polymorphs. For α- and β-$Ti_3SnC_2$, it is seen that Ti1-C bonding possesses stronger covalency than Ti2-C bonding. Thus the main contribution of hardness is come from Ti1-C bonding. The temperature and pressure dependence of bulk modulus, normalized volume, specific heats, volume thermal expansion coefficient and Debye temperature are studied fruitfully using the quasi-harmonic Debye model and the results are discussed. The bulk modulus of α-$Ti_3SnC_2$ and β-$Ti_3SnC_2$ decreases with increasing temperature. The reduction in volume for β-$Ti_3SnC_2$ is comparatively lower than that of α-$Ti_3SnC_2$ when pressure is applied. This is due to higher bulk modulus of β-$Ti_3SnC_2$. The heat capacities increase with increasing temperature, which confirms that phonon thermal softening occurs when the temperature increases. The volume thermal expansion coefficient $\alpha_V$ of α-$Ti_3SnC_2$ and β-$Ti_3SnC_2$ increases rapidly with increasing temperature at low temperature region of T < 300 K and increases gradually at high temperature region. The Debye temperature $\Theta_D$ of α- and β-Ti$_3$SnC$_2$ remains the same when $T < 100$ K and decreases linearly for $T > 100$ K.